# Irradiation tests performed on the Herschel/Pacs bolometer arrays

B. Horeau, A. Claret, L. Rodriguez, N. Billot, O. Boulade, E. Doumayrou, K. Okumura, J. Le Pennec

*Abstract*—A new concept of bolometer arrays is used for the imager of PACS, one of the three instruments aboard the future Herschel space observatory. Within the framework of PACS photometer characterization, irradiation tests were performed on a dedicated bolometer array in order to study long-term and short-term radiation effects. The main objective was to study particles impacts on the detectors applicable to future observations in orbit and possible hard and/or soft curing to restore its performances. Cobalt-60 gamma ray irradiations did not show significant degradation, so we mainly focused on single events effects (SEE). Protons and alphas irradiations were then performed at the Van de Graaf tandem accelerator at the *Institut de Physique Nucléaire* (IPN, Orsay, France), respectively at 20MeV and 30MeV. Observation showed that the shape of signal perturbations clearly depends on the location of the impacts either on the detector itself or the read-out circuit. Software curing has then to be anticipated in order to deglitch the signal. This test gives also a unique opportunity to measure some parameters of the detector: electrical crosstalk and thermo-electrical time constant. However a detailed bolometer model is necessary to understand the contribution of the thermal response in relation with the electrical response. It will be the second step of our study. Finally the complete radiation evaluation proved that this detector can be used in spatial experiments.

## I. INTRODUCTION

THE Herschel Space Observatory is the fourth of the original cornerstone missions in the European Space Agency Horizon 2000 science plan. It will be launched by an Ariane 5 rocket in the course of 2008 and will be injected in an orbit around the Sun-Earth Lagrangian point L2 for an operation time of 3.5 years. Herschel payload consists of three instruments HIFI, SPIRE and PACS devoted to spectroscopic and imaging observations in the 60μm to 670μm wavelength range. PACS covers the 60-210μm range and is both an imaging spectrometer using photo-conducting detectors, and an imager using novel technology bolometers. Herschel will study the star formation regions and also peer into the distant Universe. This will allow the reconstruction of the star formation history of the Universe during the last 10Gyr. For more detailed descriptions, see Ref.1 for the Herschel mission, and Ref. 2&3 for the PACS instrument.

Within the framework of PACS photometer evaluation we studied the behavior of the bolometer arrays in radiation environment. We have first observed the radiation effects at long-term and after at short-term. The main objective is to discriminate the future degradations of detector in orbit in order to foresee hard and soft curing to restore the initial performances. We have thus performed the Total Ionizing Dose test (TID) and high-energy particle irradiation test to study the Single Event Effects (SEE).

We then present the radiation effects and particularly the results of the high-energy particle impacts performed on a dedicated bolometer array representative of the flight model PACS imaging camera. After a short description of the bolometer arrays we briefly address the radiation environment during the lifetime of the Herschel mission. We report the results of the TID test and then describe the protons and alphas irradiations measurement performed at the Orsay Tandem accelerator and the SEE observed on this new type of bolometer. We also took advantage of the high-energy particle irradiation to observe the "pulse response" of the detector and determine the thermo-electrical time constant and checked a possible electrical crosstalk. We finally summarize the behavior of the bolometer array under high-energy particle impacts and evaluate possible consequences on the future observations in orbit.

## II. THE BOLOMETER ARRAYS

### A. Bolometer arrays description

The detection principle at work in the PACS bolometer arrays is the resonant absorption of the sub-millimetre electromagnetic radiation. In this mode, an absorption layer matched to vacuum impedance is located above a reflector. In a classical point of view, standing waves generated between incident and reflected radiations allow a theoretical thermal absorption up to 100% for a wavelength equal to four times the distance between reflector and absorber (see Fig. 1). The metal absorber is deposited on a crystalline silicon mesh insulated from a heat sink by four thin silicon rods (2μm x 5μm section - 600μm long). The time constant of the detector is given by the heat capacity of the insulated structure and the thermal conductance of the rods. A semiconductor thermometric structure fitted out on the mesh measures the temperature elevation of the sensitive part.

The PACS bolometer arrays use two silicon chips. One contains the absorbing insulated meshes (the pixels) with thermometers (Reference resistor $R_{REF}$ and bolometer resistor $R_{BOLO}$ from a tenth TΩ to a few TΩ range) and the other contains the reflectors (in gold), the cold CMOS readout electronics and the multiplexing circuit.

Working at a temperature of 300mK, the Noise Equivalent Power (NEP) of this detector approaches $2.10^{-16}$ $W.Hz^{-1/2}$.

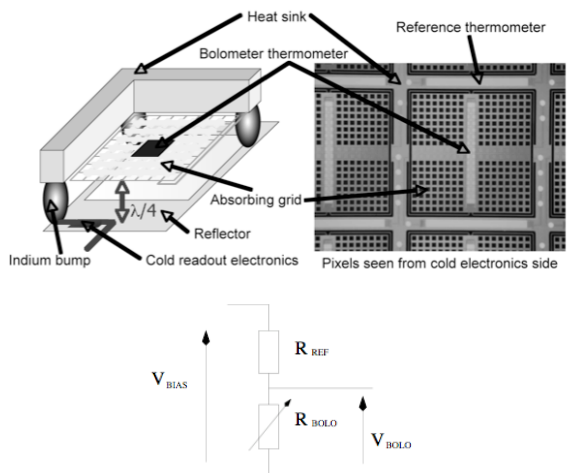

Fig. 1. The structure of the bolometer pixel and the electrical setup of the bolometric bridge for a single pixel

### B. Expected single event effects

When a particle hits the absorbing grid (5µm thick), a part of its energy is deposited. The grid is rapidly thermalized by phonons running on the entire structure. Subsequently the heat starts to flow to the cold sink through the four microscopic beams attached to the inter-pixel walls. The time constant of the detectors is then determined by the heat capacity of the grid compared to the thermal conductance of the suspension rods. As the collision brings energy to the detector, its temperature increases and its resistance ($R_{BOLO}$) decreases. As the output signal of the electric chain is the opposite of the middle point voltage we thus shall get negative glitches with a rapid rise time and a slow decay.

If the particle hits the inter-pixel wall (400 µm thick) the energy deposition is higher but the temperature elevation is smaller due to a larger heat capacity (bulkier material) warming up the detector heat sink for a short time and then lowering the resistance of the reference thermometer and moving the voltage of the bridge downwards. We shall get short positive glitches with smaller amplitude than the negative glitches.

When the particle hits the readout silicon level, the ionization phenomenon can affect for a long time the output level and gain.

### C. Space environment

The spatial environment encountered by Herschel during operations will consist of protons (~80%), alphas (~14%) and heavy ions (~4-5%) from galactic cosmic ray, solar events and solar wind plasmas. The shield around PACS can be represented by an 11 mm thick Aluminum sphere (Ref. 4), which corresponds to 3 g/cm$^2$ of Aluminum. Only proton and alpha with energy greater than 90MeV and 200MeV respectively can go through the shield. Considering the cosmic particle spectra as derived by OMERE software (Ref. 5) at L2, we expect that 0.29 particles/cm$^2$/s/sr on the PACS detectors (0.27 particles/cm$^2$/s/sr coming from protons, and 0.02 particles/cm$^2$/s/sr coming from alphas). If we note Ψ this incoming particle flux (in particles/cm$^2$/s/sr), the incident flux passing through the surface element dS is given by Ψ×π×dS.

The incoming flux passing through all faces of the detector is then Ψ×π× 2(AB+AC+BC) with A and B the length (1.2 cm long) and C the height (0.04 cm high) of the bolometer array. This leads to the expected rate of 3 particles/s for a single module, which represents the contribution of primary particles. Primary particles passing through the surrounding materials also produce about 50%-80% of additional events (Ref. 6) either by nuclear reactions or gamma-ray emissions. We then expect roughly 5 particles/s on each individual PACS module.

### III. IRRADIATIONS TEST

#### A. TID test

Considering the radiation environment during the mission a TID test was performed using a Cobalt-60 gamma ray source on a dedicated qualification model bolometer array. The purpose was to observe any damage due to protons and electrons after an irradiation level equivalent to the expected cumulated ionizing dose at the end of the mission (radiation specification of 10krads). The TID test was first performed at high dose rate (~5krad.h$^{-1}$) up to an equivalent dose of 20krad. Strong temporary degradations were observed in gain and response. Considering this rate is not representative, TID was then performed at low dose rate (~44rad.h$^{-1}$) up to a dose of 11krad. No significant degradation, either in thresholds or gains of the bolometer array has been observed. Hard curing is thus not necessary in orbit. This detector has essentially a thermal behavior and radiation perturbations are mainly expected with high-energy particles.

#### B. High-energy particle irradiation test

*1) Accelerator and detector configuration*

The particle beams diffused on a gold foil (100 or 1000nm thick) to get at the detector location a flux at least comparable to the expected space flux (particle current of a few nA).

The temperature of the detectors is maintained below 300mK by a double stage 3He/4He cryocooler. Detector response can be measured during irradiations due to a modulated source inside cryostat. A conic shield in front of the bolometer array prevents stray light to reach the detectors. All measurements were done with the blackbody source at 23K achieving ~2pW/pixel at a frame rate of 40Hz. Two kinds of acquisitions were obtained one with chopper on, to evaluate the evolution of response of the pixels, the others without chopping to record and visually analyze the particles impacts.

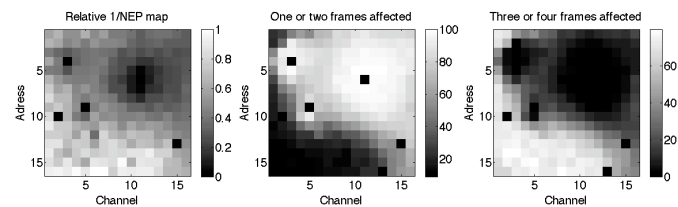

Fig. 2. On the left hand side, the inverse of the NEP (Noise/Response) map. On the centre the spatial distribution for 1 and 2 frames affected. On the right hand side, the spatial distribution for 3 and 4 frames affected (% of all glitches selected during proton irradiation). The number of frame includes the decrease and the decay of the glitch.

*2) Detector characteristics*

The bolometer array is representative of the flight model photometer. On the left hand side on figure 2 is shown the inverse of the relative NEP (noise/response) map. The most sensitive pixels are then closed to 1. We observe disparity between pixels and particularly a lower sensitive area around the pixel (12,6) and few pixels are not functional.

*3) Beam settings and glitch analysis*

We consider a glitch due to the passage of one particle when the amplitude of one spike varies over ±5 sigma from the median of the irradiation sequence (without chopping mode). The spikes can be positives or negatives and the passage of the particle can affect few frames essentially due to thermal effect. Considering one glitch (one or few frames affected) as one particle, protons fluency measurement gives ~3 particle/sec/pixel ($i_{beam}$~6nA and target thickness ~100nm) and ~0.15 particle/sec/pixel for the alphas ($i_{beam}$~20nA and target thickness ~1000nm). Nevertheless we observe a strong dispersion of the protons fluency with the expected value calculated from the Rutherford formula. We also observe a shadow due to the conic shield located in front of the detector; let's see the number of particles selected for the first pixels closed to 0. Figure 6 shows the number of particle/sec/pixel for proton (6nA) and alpha (20nA) irradiations. The higher level of noise explains the deficit of number of particle selected for the pixels from channels 14 to 16.

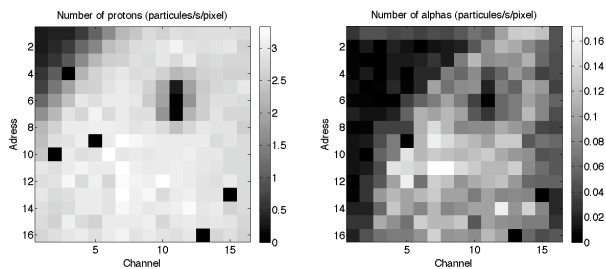

Fig. 3. Number of particle per seconds per pixel selected on the bolometer array for protons (i~6nA with 100nm thick gold) and alphas (i~20nA with 1000nm thick gold) irradiations. The black pixels are considered as dead pixels expect on top and left due to the absorption of the ions by the conic shield.

## IV. SINGLE EVENT OBSERVED

The main effects were negative glitches, positive glitches and DC level shifts. Figure 4 shows two zooms of proton irradiation sequence (one pixel) and a zoom of a typical negative glitch from alpha particle from the most sensitive region (address 12 to 16 to channel 1 to 10).

### A. The negative glitches

Most of the perturbations were negative glitches due to the passage of ions through the mesh grid (5μm thick, see on Fig. 1). The effect is mainly thermal. We observe a steep decrease (mainly one frame) and an exponential decay. The ions overheat locally the bolometer structure and the pixel (bolometer thermometer) then read the thermal perturbations. The relaxation is obtained by the link to the heat sink at the temperature of 300mK. We then analyze for proton and alpha irradiations the number of frame affected by the passage of one particle. We then select the glitches where 1, 2, 3, 4 and more frames are affected. Figure 2 shows the spatial distributions when 1 and 2 frames are affected and 3 and more frames are affected. This number of frame includes the decrease and the decay of the glitch. Both cases clearly show a spatial correlation between the ions signature (number of frame affected) and the NEP map. Two effects take place in the selection: the response and the noise of the bolometer. We expect that the more sensitive the pixel the greater the amplitude of the glitch. The length of the glitch also depends on the level of noise what explains the correlation between the NEP and the length of the glitch.

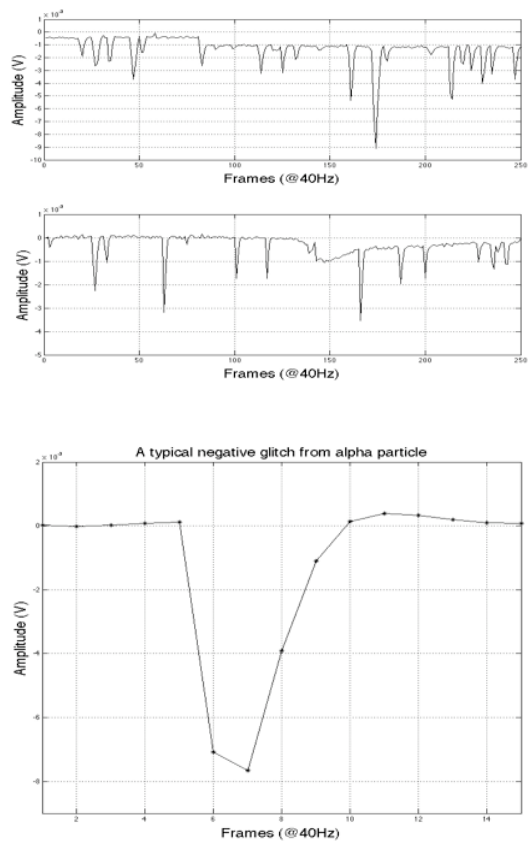

Fig. 4. On top two zooms from proton irradiation sequence with negative glitches and DC level shifts. On bottom a typical glitch from alpha particle with a steep decrease and a slow decay (glitch 4 frames long).

If we consider the most sensitive pixels with glitches 4 frames long (see Fig.6) we measure a mean amplitude of ~2.7mV and ~6.2mV respectively for the protons and the alphas. We find a factor ~2.2 whereas the energy deposit should be a factor of ~10 (measured in 5μm thick Silicon with SRIM2003, see Ref. 7).

For the less sensitive pixel (around the pixel (11,6)) the glitches are very rapid (one frame only) with amplitudes of few mV. We also notice overshoot for the heaviest particle but a detailed bolometer model is necessary to go further into the analysis.

Chopping mode acquisitions were also performed with an optical modulation (~23K) at 0.7Hz. The difference between a reference acquisition (no irradiation) and an irradiated acquisition doesn't show variation, which means no gain variation.

### B. The DC level shifts

The DC level shifts (see proton sequence zoom on Fig. 4)

can be explained by the passage of the particle through the MOS electronics located just below the detector circuit. The shifts can be positives or negatives. This is a rare event (only observed with high flux of protons) that affects the baseline of few mV for a long time (several tens of seconds). Concerning the positive shifts, the mean amplitude is 1.6mV (±0.59mV) from the median.

*C. The positive glitches*

We observe as well rapid positives glitches (one frame affected only) due to the ions interactions with the inter-pixel wall. The heat sink heating can then affect several contiguous pixels mostly observed with alphas (~2.3 pixels on average for an alpha particle). Indeed the alpha ion (at 30MeV) is almost totally absorbed in the inter-pixel wall 400µm thick. We then select contiguous pixels (around the maximum amplitude value) "reached" by one particle at the same time. Figure 5 shows a bolometer array image with 7 impacts (at different time). That shows different shapes of pixels affected.

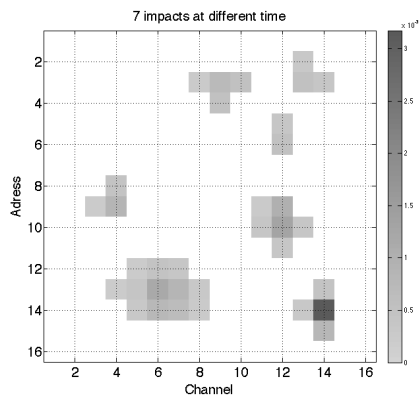

Fig. 5. An image of the bolometer array with 7 alpha impacts at different time. The amplitude is in Volts.

V. INTERPRETATION AND RESULTS

First conclusion is that the bolometer signal is not significantly affected. Particles affect only few frames and the perturbations can then be removed by a standard deglitching method. There is also not effect on the gain of the detector. High-energy particle irradiation can be particularly relevant to observe electrical crosstalk. Given the probability to have two particles at the same time in the entire matrix, there is no evidence that particle hits affect several contiguous pixels.

We can also deduce from the amplitude vs. time diagram the time constant of the detector despite a low sampling rate. Although the heat capacity and thermal conductance are very dependant on the temperature, a mean time constant can be deduced from these measurements. Figure 6 shows these diagrams for protons and alphas for the most sensitive pixels (address 12 to 16 to channel 1 to 10) and the glitches 4 frames long. We note that the higher the amplitude, the higher the bandwidth. If we consider $\tau \sim 24ms$ we then obtain a typical bandwidth of $B=1/(2\pi\tau)\sim 6Hz$. The thermo-electrical bandwidth is equivalent to the optical bandwidth measured during flight model calibration (see Ref. 3).

VI. CONCLUSIONS

Concerning the total ionizing dose, the PACS bolometer arrays can withstand the spatial environment at L2 without hard curing. We then used proton and alpha beams in order to observe the signal perturbations of high-energy particle impacts. The main SEE's are positive glitches (due to inter-pixel wall hits) and negative glitches (due to absorbing grid hits). We did not observe any gain variations and any electric crosstalk. Only inter-pixel wall impacts can affect few pixels. In both cases SEE doesn't affect in a significant way and deglitching method can be easily implemented to remove the perturbations during the observations in orbit. We also determined the thermo-electrical bandwidth (~6Hz) that is the same order of magnitude of optical bandwidth. However a detailed bolometer model is necessary to understand the contribution of the thermal response in relation with the electrical response. It will be the second step of our study.

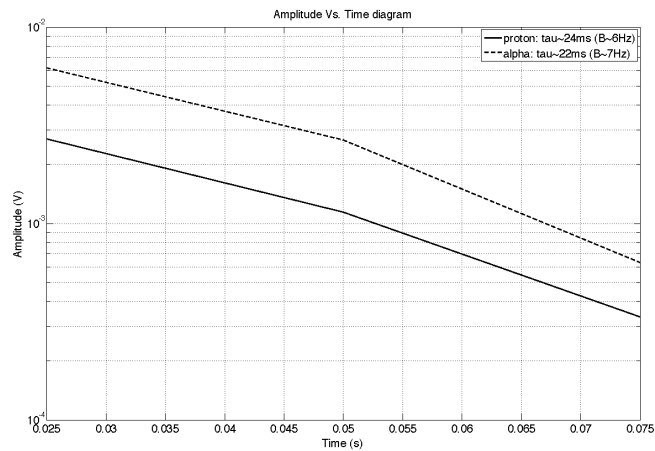

Fig. 6. Amplitude vs. time diagram for the most sensitive pixels (address 12 to 16 to channel 1 to 10) and glitches 4 frames long. *B* is the thermo-electrical bandwidth and *tau* the time constant deduced from the diagram.

ACKNOWLEDGMENT

The authors wish to thank the team at IPN and Bernard Rattoni from CEA/DIMRI/SIAR for their help during the irradiation tests.